\documentclass[proof]{WileyASNA-v1}

\articletype{Article Type}%
\usepackage{color}

\received{8 September 2019}
\revised{\today}
\accepted{XX YY 2019}

\begin{document}

\title{SED modelling of broadband emission in the pulsar wind nebula 3C~58}

\author[1]{Seungjong Kim}

\author[1]{Hongjun An*}

\authormark{Kim and An}

\address[1]{\orgdiv{Department of Astronomy and Space Science}, \orgname{Chungbuk National University}, \orgaddress{\state{Cheongju-si 28644}, \country{Republic of Korea}}}

\corres{*Hongjun An, \email{hjan@cbnu.ac.kr}}

\presentaddress{Chungdae-ro 1, Seowon-gu, Cheongju 28644, Republic of Korea}

\abstract{We investigate broadband emission properties of the
pulsar wind nebula (PWN) 3C~58 using a spectral energy distribution (SED) model.
We attempt to match simultaneously the broadband SED and spatial variations of
X-ray emission in the PWN. We further the model to explain a possible far-IR
feature of which a hint is recently suggested in 3C~58: a small bump at $\sim$$10^{11}$\,GHz in the
{\it PLANCK} and {\it Herschel} band.
While external dust emission may easily explain the observed bump,
it may be internal emission of the source implying an additional population of particles.
Although significance for the bump is not high,
here we explore possible origins of the IR bump using the emission model and
find that a population of electrons with GeV energies can explain the bump.
If it is produced in the PWN, it may provide new insights into
particle acceleration and flows in PWNe.}



\maketitle

\section{Introduction}\label{sec1}
	A pulsar wind nebula (PWN) is a remnant of supernova explosion of a massive star
and is powered by an energetic central pulsar. It is believed that the pulsar's
relativistic (cold) plasma wind interacts with ambient medium, forming
termination shock \citep[][]{Kennel_1984}. The shock then accelerates
cold pulsar-wind particles, and the accelerated particles and $B$ flow outwards to
form a PWN. So PWNe have characteristic morphology having
a central pulsar, a torus corresponding to the termination shock, polar jet outflows,
and an extended nebula. These structures are best seen in the X-ray band,
and an archetype of PWNe is the Crab nebula \citep[][]{Weisskopf_2000, Madsen_2015}.
Although the detailed morphologies of PNWe are complex, recently
magnetohydrodynamic (MHD) simulations \citep[e.g.,][]{Komissarov_2004}
were able to reproduce the basic structure of the Crab nebula.

	Interaction of a pulsar's wind and the ambient medium can be various, and so
different types of PWNe are observed.
The characteristic torus-jet structure may be crushed if the pulsar moves fast,
and then a bow shock and a long tail may be formed \citep[][]{Cordes_1993}.
Intrabinary shock produced by interaction of pulsar and stellar winds in pulsar binaries
is also a type of PWNe. Flows and emission in
various PWNe share the same fundamental physics but with different geometrical effects
\citep[][]{Romani_1997, Dubus_2006, Romani_2016, An_2017}.
With the observational diversity in different types of objects,
PWNe are very useful to study physics of relativistic shock acceleration
and astrophysical plasma flow \citep[see][for reference]{Gaensler_2006,Reynolds_2017,Kargaltsev_2017}.
This can be done by modelling the emission spectra of PWNe which are well characterized
by double-hump structure: a low-energy hump produced by synchrotron radiation of electrons and
a high-energy hump by inverse-Compton (IC) upscattering of soft-photon fields by the
energetic plasma particles.

	Extended PWNe may exhibit spatial variations in their emission properties,
and these have been investigated observationally for some bright PWNe in the X-ray band
\citep[e.g., G21.5$-$0.9 and MSH~15$-$5{\sl 2};][]{Nynka_2014,An_2014}.
However, current SED models are developed mainly to explain spatially-integrated emission, and
models for spatial variation focus on narrow-band properties \citep[e.g., the X-ray band;][]{Tang_2012,Porth_2016}.
Since all the observed properties need to be put together for better understanding of PWNe,
it is crucial to have a model that can explain the spatially-varying multiband properties
simultaneously.

	The pulsar wind nebula 3C~58 is an X-ray bright object with
clear torus-jet structure (Fig.~\ref{fig:fig1}). The PWN is large \citep[$10'\times6'$ corresponding
to $R_{\rm pwn}\approx 3.7$\,pc for an assumed distance of 3.2\,kpc;][]{Roberts_1993}
and was suggested to be possibly associated
with SN~1181 \citep[e.g.,][]{Stephenson_1971}
implying an age of $\sim$800\,yr; this association is controversial though \citep[e.g.,][]{Bietenholz_2001}.
As the PWN is bright across electromagnetic wavebands, a high-quality SED and
spatial variations of the emission were measured well
\citep[][]{Slane_2004,Slane_2008,Aleksic_2014,Abdo_2013,Ackermann_2013,Li_2018}.
Recently, a possible spectral cutoff
at $\sim$25\,keV (An 2019) and a small bump at $\sim$$10^{11}$\,GHz \citep[][]{Planck_2016,kpa19}
were also suggested. In particular, the latter may imply that there may be two populations
of accelerated electrons in the PWN.
A similar SED feature was also seen in the Crab nebula
and was attributed to emission of an additional population of electrons in the PWN
\citep[][]{Bandiera_2002} although the existence of the bump
is controversial in this source \citep[e.g.,][]{Macias_2010}.

	In this paper, we investigate the possible IR bump in 3C~58 using an SED model.
In Section~\ref{sec2} we summarize observational features and
previous modelling efforts. We present our SED model for PWN emission and results of modelling
in Sections~\ref{sec3} and \ref{sec4}. We then discuss the results and conclude in Section~\ref{sec5}.
We assume a distance of 3.2\,kpc to 3C~58.

\section{Observational properties of 3C~58 and previous modelling}\label{sec2}
\subsection{Observational properties}\label{sec2_1}

\begin{figure}[t]
\centerline{\includegraphics[width=78mm]{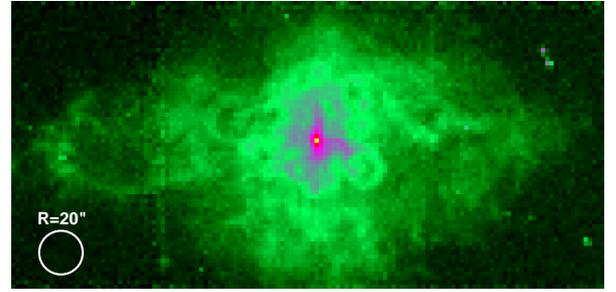}}
\vspace{-2 mm}
\caption{A {\it Chandra} image of the PWN 3C~58. An $R=20''$ circle is shown in the lower left corner for reference.
\label{fig:fig1}}
\vspace{-4 mm}
\end{figure}

\begin{figure*}[t]
\begin{tabular}{cc}
\includegraphics[width=85mm]{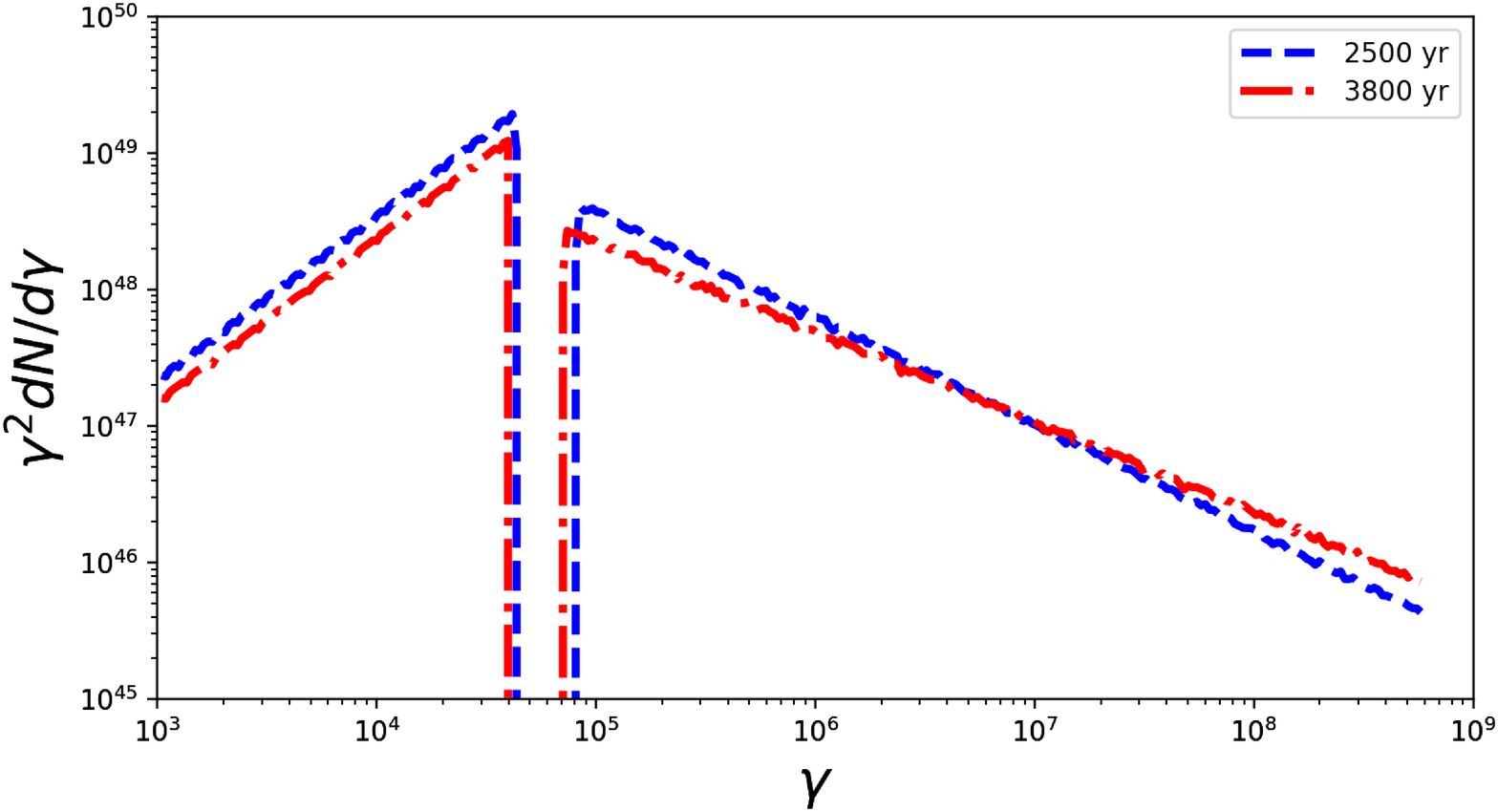} &
\includegraphics[width=85mm]{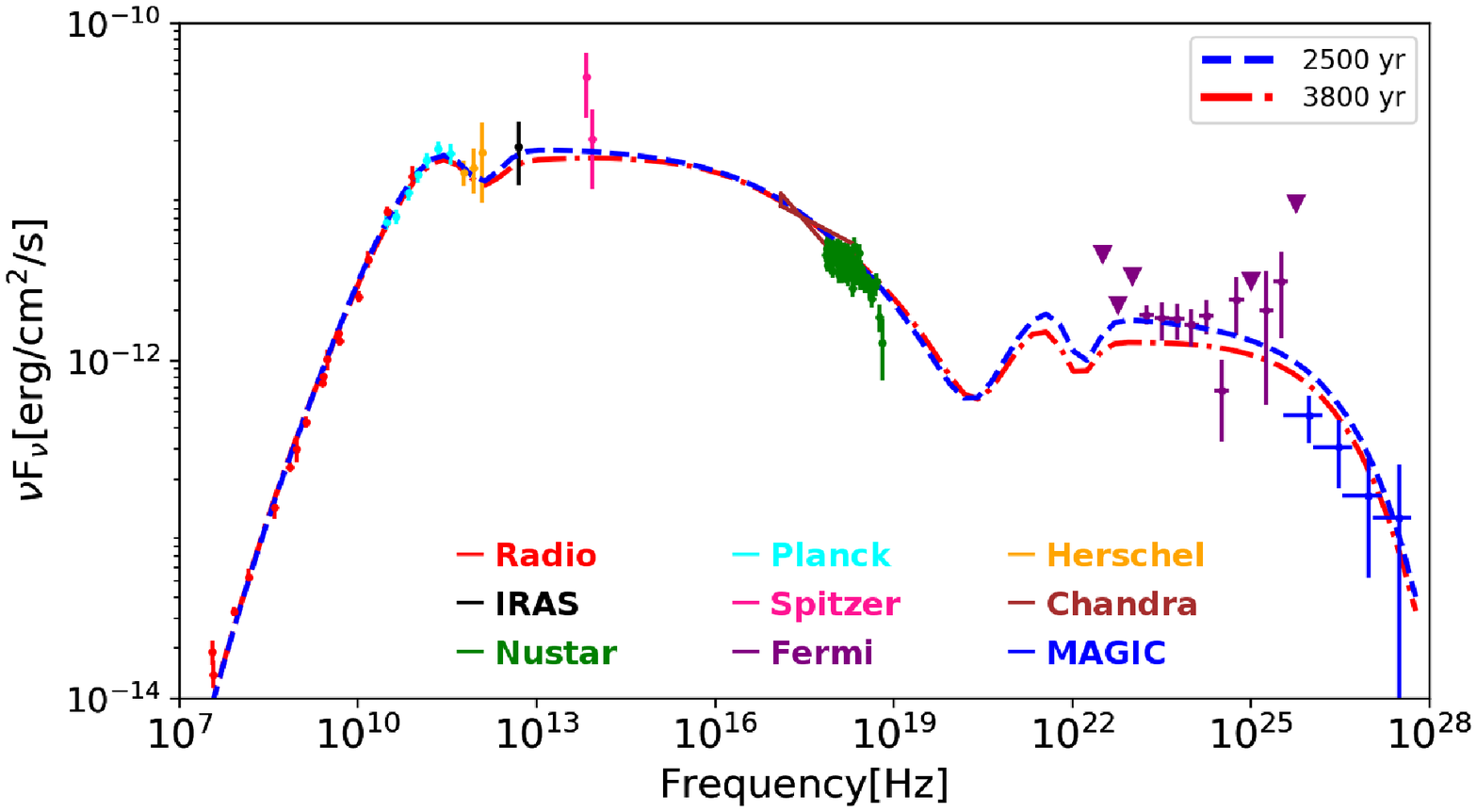} \\
\vspace{-2 mm}
\includegraphics[width=85mm]{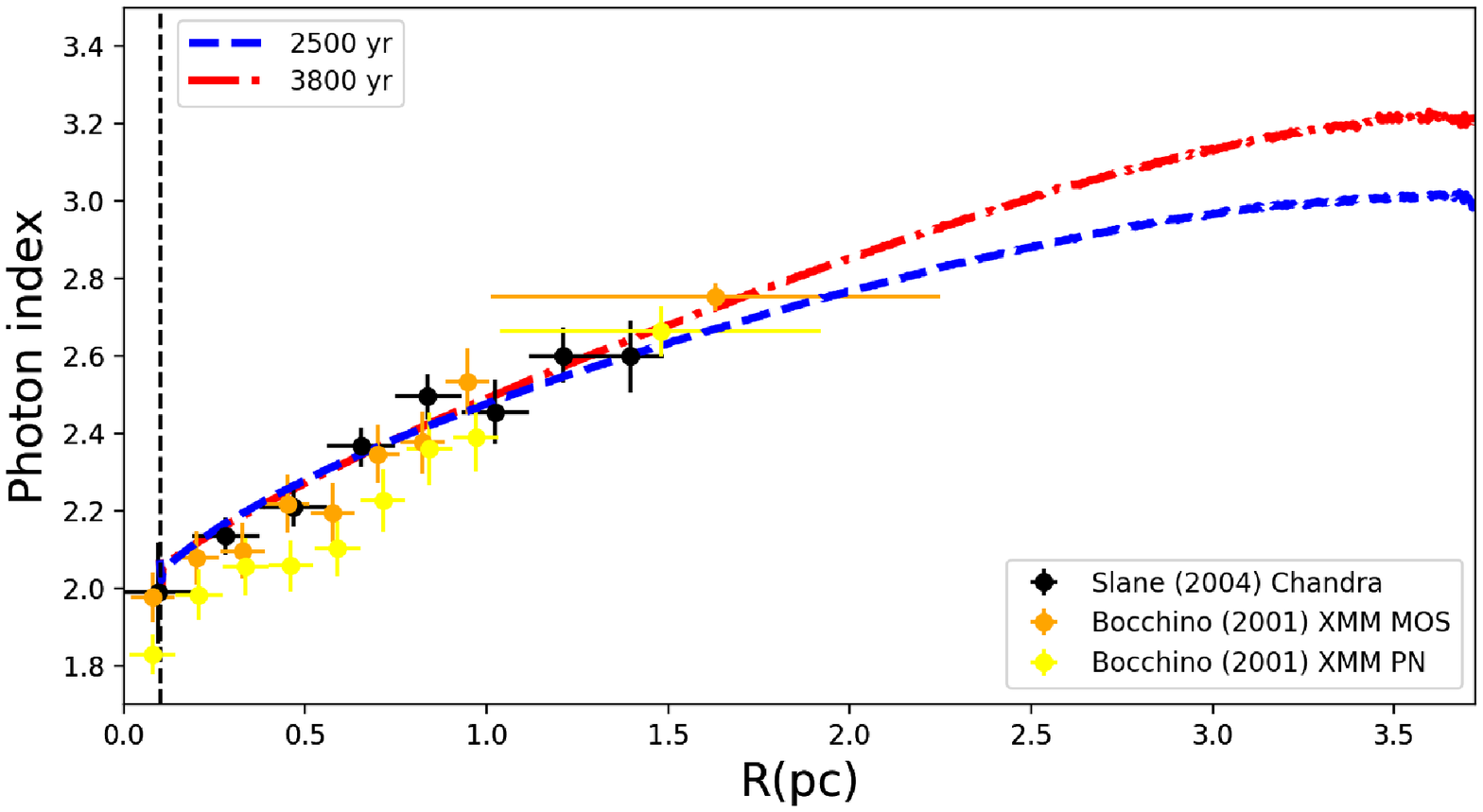} &
\includegraphics[width=85mm]{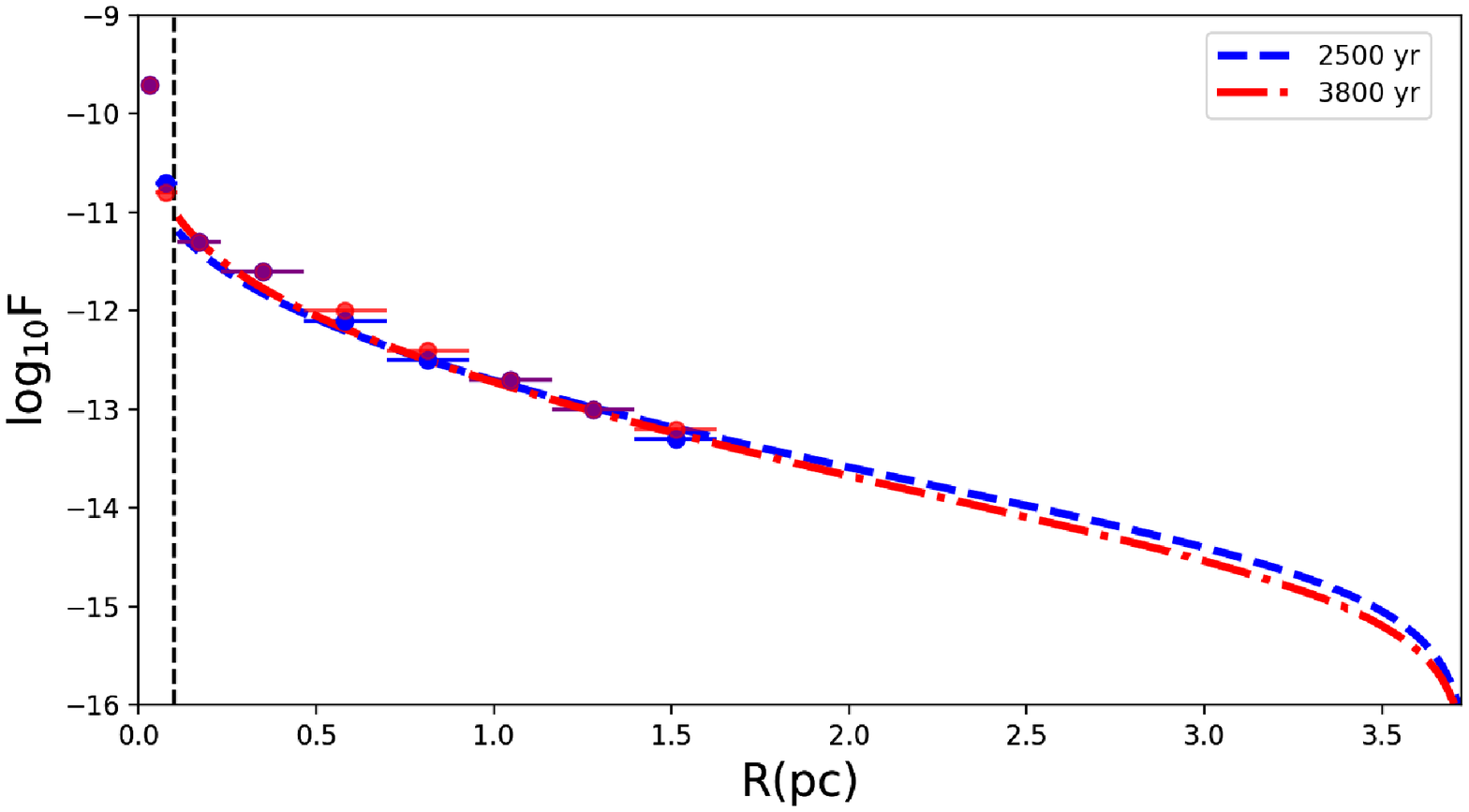} \\
\end{tabular}
\vspace{-4 mm}
\caption{Observational data (taken from literatures; see text)
and our models of 3C~58 emission (for two different ages).
{\it Top left}: injected electron distributions. {\it Top right}: a broadband SED.
{\it Bottom left}: X-ray photon-index profiles. {\it Bottom right}: the 2.2--8\,keV surface-brightness profile.
Note that small differences in the radial profiles of the photon index (bottom left)
measured with {\it Chandra} and {\it XMM-Newton} are because
of different bands used \citep[see][]{Slane_2004, Bocchino_2001}:
2.2--8\,keV for {\it Chandra} and 0.5--5\,keV for {\it XMM-Newton}.
Vertical lines show the injection site ($R_{\rm inj}=$0.1\,pc).
\label{fig:fig2}}
\vspace{-3 mm}
\end{figure*}

	As 3C~58 is bright in the broad waveband and has an important pulsar in it \citep[][]{Slane_2004},
the PWN was intensively studied in the past (Figs.~\ref{fig:fig1} and \ref{fig:fig2}).
Observations in the radio to X-ray band with sufficient angular resolution
revealed that the PWN has a similar morphology in the bands
but is relatively larger $10'\times6'$ in the radio band
than in the X-ray band ($\approx 8'\times5'$; Fig.~\ref{fig:fig1})
due to the synchrotron burn-off effect.
As the source is large compared to angular resolutions of current
X-ray observatories, spatial variations of the X-ray spectrum were well measured
with {\it Chandra}, {\it XMM-Newton} and {\it NuSTAR} \citep[][]{Bocchino_2001, Slane_2004, An_2019}.
The X-ray photon-index profile shows an increasing trend from the center
outwards (Fig.~\ref{fig:fig2} bottom left). At large distances, the profile is suggested to be flat,
with a break at $R\sim 80''$ perhaps due to effects of particle diffusion \citep[][]{Tang_2012},
but the break is not very clear because of possible contamination of thermal emission \citep[][]{Bocchino_2001}
and paucity of counts. The surface brightness decreases monotonically from the center
(Fig.~\ref{fig:fig2} bottom right).

	The SEDs were well sampled from the radio to TeV band (Fig.~\ref{fig:fig2} top right).
The radio SED is a simple power law with an energy index of $\alpha_{\rm r}\approx0.1$ ($F_\nu\propto \nu^{-\alpha_r}$)
up to the {\it PLANCK} band \citep[e.g.,][]{Planck_2016,Green_1992_radio_index_M}.
At around $10^{11}$\,GHz, the spectrum breaks to
an $\alpha_{\rm IR}\approx1$ power law in the IR band \citep[][]{Slane_2008}. The flat IR SED
extends to the optical band $\sim$$10^{14-15}$\,GHz, and breaks to an $\alpha_{\rm X}\approx1.3$ power law
in the X-ray band which may cut off at $\sim$25\,keV \citep[][]{An_2019}.
The break in the optical band is certainly a synchrotron-cooling break
and implies the magnetic-field strength in the source to be 30--200\,$\mu$G
for an assumed age range of 800--5400\,yr, and the possible X-ray cutoff
corresponds to the maximum electron energy of $\sim$100\,TeV.
The gamma-ray SED (Fig.~\ref{fig:fig2}) is relatively poorly characterized, but is
flat in the {\it Fermi}-LAT band \citep[GeV;][]{Li_2018}
and curves down in the TeV band \citep[][]{Aleksic_2014}.
Note that a hint of an IR bump is seen recently \citep[][]{Planck_2016, kpa19} in the
{\it PLANCK} and {\it Herschel} data (Fig.~\ref{fig:fig2});
the significance is not high because of possible contamination of Galactic foreground emission.

	Note that the data in Figure~\ref{fig:fig2} are taken from literature referred to above.
For the X-ray data, we take the 2.2--8\,keV band \citep[][]{Slane_2004} as our baseline because
this band is less affected by Galactic absorption. A photon-index profile in this band
was reported previously \citep[][]{Slane_2004}, and we reanalyzed archival {\it Chandra}
data (Obs. IDs 3832 and 4382) to generate 2.2--8\,keV surface-brightness profiles \citep[][]{An_2019}.

\subsection{Previous SED modelling}\label{sec2_1}
	With the high-quality broadband measurements, SED models were applied to the data
to infer physical properties of the plasma flow in the source \citep[e.g.,][]{Tanaka_2013,Torres_2013,Li_2018}.
These models assume stationary one-zone or spatially varying multi-zone emission, and compute
SEDs that match the observed (spatially-integrated) one.
The detailed model components and prescriptions differ among the models
but in general they all can explain the spatially-integrated SED data with reasonable magnetic-field strengths
($B$=20--80$\mu$G) and ages (1000--5000\,yr). However, the far-IR band is not very well modelled
because {\it PLANCK} and {\it Herschel} measurements lacked at the times, and the high-energy tail
of the model SEDs extends to MeV band without a cutoff, potentially conflicting
with the hint of a 25-keV cutoff. Furthermore, these models did not attempt to explain
spatial variations in the X-ray band.

	While carefully adjusting the broadband-SED model parameters may allow matches to the spatial
variations, they were modelled with slightly different approaches using X-ray data only.
A semi-analytic and a numerical diffusion models were used to explain the spatial variations
of the source's emission properties \citep[][]{Tang_2012,Porth_2016},
and the models were able to match the size and
the photon-index profiles with a diffusion coefficient of $10^{26-27}\ \rm cm^2\ s^{-1}$.
However, these diffusion models are rather limited to the X-ray band and an attempt to explain
the broadband SED of 3C~58 simultaneously was not made.

	Recently, \citet{Ishizaki_2018} tried to explain both SED and
spatial variations measured for 3C~58 by approximately solving fluid equations with 
diffusion. This model seems to explain the SED (without the possible IR feature), but
matches to the spatial variations are rather poor.

\section{The SED model used in this work}\label{sec3}
	As both broadband SEDs and spatial variations can provide important information
on plasma flow properties in PWNe, it is important to model them simultaneously.
So we develop an SED model \citep[][]{kpa19}
and attempt to explain both broadband SEDs and spatial variations of X-ray spectra in PWNe.
The model assumes a power-law (or a broken power-law) distribution of injected electrons
$dN/d\gamma_e = K_e (\gamma_e/\gamma_0)^p$, where $N$ is the number of electrons
and $\gamma_e$ is the Lorentz factor, and $\gamma_0$ is a reference point.
For spatially-varying flow properties, we use power-law prescriptions \citep[e.g.,][]{Reynolds_2009}:
the bulk flow speed $V(r)=V_0 (r/R_0)^{\alpha_V}$,
magnetic-field strength $B(r)=B_0 (r/R_0)^{\alpha_B}$, and
diffusion coefficient $D(B,\gamma_e)=D_0 (B/B_0)^{-1} (\gamma_e/\gamma_D)^{\alpha_D}$,
where $r$ and $R_0$ ($\approx 6''$; Fig.~\ref{fig:fig1}) are the distances to the
emitting zone and the termination shock from the central pulsar, respectively.

	In our initial study \citep[][]{kpa19}, we applied the model to 3C~58 focusing on
the spatially-integrated SED without considering spatial variations of the emission.
In that work, we assumed 
toroidal-magnetic flux conservation
\citep[i.e., $\alpha_V + \alpha_B=-1$;][]{Reynolds_2009} which together with a radio-expansion
speed measurement \citep[][]{Bietenholz_2006_speed} and an assumed age
constrain the model parameters $V_0$, $\alpha_V$, and $\alpha_B$.
We further assumed Bohm diffusion (i.e., $\alpha_D=1$).
By injecting one population of electrons with or without
a spectral break, we were able to match the broadband SED of 3C~58.
In that study, we find that a reasonable IR match is achieved with
a small- (2900\,yr) or large-age model (5400\,yr)
but a middle-age (3800\,yr) one underpredicts the far-IR SED significantly.
Nevertheless, these models do not match the radial profiles of the X-ray photon index and
surface brightness, and cannot explain the IR bump well unless
an external source for narrow emission (e.g., blackbody) is assumed.

	Simply adjusting the parameters of this model to explain the far-IR bump
and spatial variations of X-ray spectrum poses two problems:
\begin{enumerate}[(1)]
\item The far-IR bump is very narrow, so simply modifying the
single-population power-law (or broken power-law) distribution is not sufficient to explain
the bump. This is because electrons in PWNe undergo adiabatic and radiative cooling, and
so the emission spectra blur significantly during the flow.
\item Many of the parameters are already constrained by the model assumptions
(e.g., $V_0$, $\alpha_V$, $\alpha_B$, and $\alpha_D$), and adjusting only the rest
parameters do not significantly improve matches to the spatial variations.
\end{enumerate}
So we relax these assumptions here.

	For (1), we use two populations of electrons as suggested
observationally \citep[][]{Bandiera_2002,Meyer_2010_Crab_2_popul}
and theoretically \citep[e.g.,][]{Lyutikov_2019}.
In the latter, the authors hypothesized possible
existence of two populations of electrons \citep[][]{Sironi_2011,Lyutikov_2019}:
a low-energy distribution ($\gamma_{e}<10^{5}$; perhaps the unshocked pulsar wind)
and a shock-accelerated one with $\gamma_{e}\ge 10^{5}$. The former may be further accelerated via
turbulent reconnection and develop a high-energy tail.

	We also relax assumptions for the flow parameters (2) as the toroidal-magnetic flux conservation
and Bohm diffusion may not be strictly valid in PWNe \citep[e.g.,][]{Tang_2012, Reynolds_2009};
particle generation by filament evaporation and magnetic amplification/reconnection in PWNe are
theoretically predicted \citep[][]{Lyutikov_2003_M}.
Indirect hints of these were suggested observationally in some X-ray bright PWNe
\citep[e.g., G21.5$-$0.9 and MSH~15$-$5{\sl 2};][]{Nynka_2014,An_2014}.
	
\section{Results of modelling}\label{sec4}
	Although we relax some of the model assumptions,
the observationally-constrained flow-speed parameters ($V_0$ and $\alpha_V$) are still
set by the radio-expansion speed measurement and an assumed age as was done in our previous work.
There are still many parameters to adjust (e.g., Table~\ref{ta:ta1}), but not all of them are free;
spectral indices of electron distributions ($p$) are tightly constrained by observed slopes of
the radio and IR SEDs, and the injection sites $R_{\rm inj}$ by the X-ray image of
the inner torus \citep[Fig~\ref{fig:fig1}; see also][]{Slane_2004}.
With these constraints, we vary the other adjustable parameters to match simultaneously
the broadband SED and spatial variations, and present the results
in Figure~\ref{fig:fig2} and the parameters in Table~\ref{ta:ta1}.
As seen in the Figure, the models can explain the broadband SED,
the radial profiles of the X-ray spectrum and surface brightness of the source.

	The parameters are only slightly different
from those used in our previous modelling \citep[][]{kpa19}.
In general, $\alpha_V$ is related to adiabatic cooling which is dominant
for electrons emitting at far-IR frequencies;
a smaller value reduces the cooling ($\propto \alpha_V+2$) of the electrons.
Hence, smaller $\alpha_V$ helps to minimize SED blurring
of the far-IR bump. Second, $\alpha_B$ is related to
synchrotron cooling of IR-to-X-ray emitting electrons;
for larger $\alpha_B$, the cooling is relatively weaker in the inner regions,
and therefore the X-ray emission extends to larger distances.
The photon-index profile is mainly controlled
by $\alpha_D$ and $\gamma_{e, max}$, and we find $\alpha_D\approx0.3$ which
is similar to the result of a previous diffusion model \citep[][]{Tang_2012},
and $\gamma_{e, max}=6\times 10^{8}$ does not conflict significantly with the possible 25-keV spectral cutoff.
Since the X-ray cutoff is not yet very significant, a larger value of $\gamma_{e, max}$ may be
used; the model parameters change only slightly in this case.
Of course, actual determination of the
parameters is much more complex because the parameters covary,
and the model has to match the broadband SED and the spatial variation simultaneously.
Nevertheless, we find that the parameters in Table~\ref{ta:ta1}
are physically
plausible \citep[see ][for previous SED and diffusion-model estimates]{Torres_2013, Tang_2012}

	The inferred distributions of the two populations are power laws
in $\gamma_e=1-4\times 10^{4}$ with $p=-0.8$ and in $\gamma_e\approx 10^{5}-6\times 10^{8}$
with $p\approx-2.7$ (Fig.~\ref{fig:fig2} top left).
Note that the lower bound for $\gamma_e$ of the low-energy population is
not well constrained, and using a larger value \citep[e.g., 100;][]{Lyutikov_2019} is also possible.
A small gap (i.e., $\gamma_e=4\times 10^4-10^{5}$)
between the energy distributions is necessary
like in the case of the Crab nebula \citep[e.g.,][]{Bandiera_2002,Meyer_2010_Crab_2_popul}.
This is to reproduce the small deficit at $\sim 10^{12}$\,Hz
(i.e., the $10^{11}$\,Hz bump; Fig.~\ref{fig:fig2}); without the gap, the dip in the SED is washed
out. The low-energy distribution may correspond to unshocked polar wind with $<\gamma_e> \approx \gamma_w$
and the high-energy one to the shock-accelerated equatorial wind with
$<\gamma_e>\approx \gamma_w \sigma_w$,
where $\sigma_w$ is the magnetization parameter (magnetic-to-particle energy ratio)
and $\gamma_w$ is the pre-shock Lorentz factor of the pulsar wind.
These distributions are similar to those predicted in a theoretical model \citep[][]{Lyutikov_2019} and/or
particle-in-cell (PIC) simulations \citep[][]{Sironi_2011}.
The low-energy population is expected to have a sharp cutoff at $\gamma_e\approx \gamma_w \sigma_w$
\citep[e.g.,][]{Werner_2016,Lyutikov_2019}, but \citet{Lyutikov_2019} hypothesized that
the distribution may extend to higher energies via turbulence acceleration.
So we search for a high-energy tail in the low-energy population with our SED model,
but find that the high-energy cutoff of
the low-energy distribution needs to be sharp (e.g, a power law with a slope $p_2\le-5$)
for 3C~58. Otherwise, it is hard to explain the IR bump with the model.
We note that two populations are used only to match the far-IR bump;
the model can explain the SED and spatial variations simultaneously
with one population if we ignore the possible far-IR bump.

\begin{center}
\begin{table}[t]%
\centering
\caption{ Parameters of the SED models \label{ta:ta1}}%
\vspace{-2 mm}
\tabcolsep=0pt%
\begin{tabular*}{20pc}{@{\extracolsep\fill}lcc@{\extracolsep\fill}}
\toprule
\textbf{Parameter} & \textbf{2500 yr} & \textbf{3800 yr}   \\
\midrule
$B_0$ ($\mu$G) & $140$ & $140$\\
$\alpha_B$ & $-0.35$ & $-0.3$\\
$V_0$ & $0.23c$ & $0.01c$\\
$\alpha_V$ & $-1.3$ & $-0.5$\\
$D_0$ ($\rm \ cm^{2}\ s^{-1}$) & $1.7\times 10^{27}$ & $1.7\times 10^{27}$\\
$\alpha_D$ & 0.3 & 0.3\\
\multicolumn{2}{l}{Low-energy population$^\dagger$:}  \\
$p$ & $-0.8$ & $-0.8$ \\
$\gamma_{e, min}$ & $1$ & $1$\\
$\gamma_{e, max}$ & $4\times 10^4$ & $4\times 10^4$\\
\multicolumn{2}{l}{High-energy population$^\dagger$:} \\
$p$ & $-2.78$ & $-2.66$ \\
$\gamma_{e,min}$ & $8\times 10^4$ & $7\times 10^4$\\
$\gamma_{e,max}$ & $6\times 10^8$ & $6\times 10^8$\\
\multicolumn{2}{l}{Soft-photon field for IC$^\ddagger$:}  \\
Temperature (K) & 20 & 20\\
Energy density ($\rm eV/cm^{3}$) & 5 & 6\\
\bottomrule
\end{tabular*}
\begin{tablenotes}
{\footnotesize \item[$^\dagger$] Injected at the termination shock $R_{\rm inj}=0.1$\,pc
\item[$^\ddagger$] Although we consider self-Compton, and IC of CMB and Galactic IR fields
for the gamma-ray SED, we show only the Galactic IR background here
because contribution of the others is very small}
\end{tablenotes}
\vspace{-3 mm}
\end{table}
\end{center}

\section{Discussion and Conclusions}\label{sec5}
	We modelled the broadband SED and spatial variations of the X-ray emission properties of 3C~58 using
an SED  model. The model with physically plausible parameters could explain
the broadband SED and the spatial variations simultaneously.
We then investigated the possible far-IR SED bump in the PWN using
the same model and found that an additional population of electrons is needed in the
$\gamma_e=1-4\times 10^4$ range although the lower bound is not very certain.

	From the modelling, we find that small-age models (e.g., 800\,yr) are hard to accommodate
the expansion speed \citep[e.g.,][]{Bietenholz_2001}, and large-age models
\citep[e.g., 5400\, yr of the pulsar's characteristic age;][]{Livingstone_2009}
are unlikely as the far-IR bump blurs significantly
and gamma-ray emission is too strong to match the LAT upper limits.
So the age of 3C~58 is constrained,
and its association with SN~1181 is unlikely in our model.

	By simultaneously matching the SED and spatial variations,
we infer plasma flow properties in 3C~58. Our results generally agree with previous
ones, but our model predicts that flattening in the X-ray photon-index profile occurs
at a larger radius $R\approx 3'$ (Fig.~\ref{fig:fig2}) than $R\approx 1.5'$
in a diffusion model of \citet{Tang_2012}. {\it XMM-newton} data seem to support
our model (Fig.~\ref{fig:fig2}), but the data are in a different energy band (0.5--5\,keV).
So further X-ray studies are warranted.

	Intriguingly, we find that it is hard to explain the observations if we
require toroidal magnetic-flux conservation; the bulk flow speed and/or the magnetic-field need
to drop more quickly for the 2500-yr model or slowly for the 3800-yr one
than for the conserved cases \citep[][]{kpa19}.
This may suggest that the kinetic energy may be converted into other forms (e.g., turbulence) and
the magnetic field dissipates/amplifies in the PWN, perhaps by turbulent reconnection.
The unshocked polar wind may be accelerated at these sites. Then, we may inject the
low-energy population at different locations (e.g., $R_{\rm inj}\ge 0.1$\,pc)
than at the termination shock.
This will make it easier to match the far-IR bump with the model
because blurring of the bump is less a concern if $R_{\rm inj}\ge 0.1$\,pc.

	The far-IR feature seen in 3C~58 is not very significant,
and so the bump could be just statistical fluctuation of the measurements.
Or the feature may be produced by external dust emission ($T_{\rm dust}$$\approx$10\,K);
studies of the Galactic dust-temperature distribution show that
low-$T$ ($T_{\rm dust}$$\ge$10\,K) regions exist \citep[e.g.,][]{zh14,PLANCKdust16}.
These can explain the far-IR feature in 3C~58.
Alternatively, a similar bump also seen in the Crab nebula \citep[][]{Bandiera_2002}
and a theoretical prediction \citep[][]{Lyutikov_2019} may suggest
that far-IR bumps may be produced in PWNe. Hence, we investigated this possibility.
While our modelling of two populations in 3C~58 is in line with the internal emission scenario,
we note that detailed shapes of the electron distributions differ from
theoretical ones: a broad Maxwellian-like one (low energy) little affected by shock
and a shock-accelerated one (high energy). In particular, the low-energy distribution
we inferred (Fig.~\ref{fig:fig2}) does not appear to be Maxwellian-like.
We also checked to see if the low-energy population has a hard high-energy
tail as hypothesized previously \citep[][]{Lyutikov_2019}.
Our model prefers a sharp cutoff at $\gamma_e=4\times 10^{4}$, implying no
significant high-energy tail. However, it is still possible that a weak tail
indiscernible with the current data exists in the distribution.
The conclusions we draw here about the low-energy particle distribution
are not very strong since we are assuming that the weakly-detected far-IR hump
is produced in the PWN. Nevertheless, these findings, if real, 
may provide new insights into PWNe physics and particle acceleration
in relativistic shocks.

	It is crucial to detect the far-IR bump clearly and tell conclusively whether
the bump is external or internal.
This can be done with deep far-IR observations, but Galactic foreground emission
in that band may preclude a
firm detection. An alternative way is to observe 3C~58 in the MeV band.
In particular, our model predicts a corresponding gamma-ray bump
at $\sim$10--100\,MeV (Fig.~\ref{fig:fig2}). This can be tested with near-future gamma-ray
observatories \citep[e.g., AMEGO, e-ASTROGAM;][]{McEnery_2017_M,eastrogam}.

\section*{Acknowledgments}
We thank the referee for helpful comments.
This research was supported by Basic Science Research Program through
the \fundingAgency{National Research Foundation of Korea (NRF)}
funded by the \fundingAgency{Ministry of Science, ICT \& Future Planning} (\fundingNumber{NRF-2017R1C1B2004566}).

\bibliography{3C58_han}%

\end{document}